\def\lesssim{\,\lower2truept\hbox{${<\atop\hbox{\raise4truept\hbox{$\
sim$}}}$}\,}
\def\gtrsim{\,\lower2truept\hbox{${>
\atop\hbox{\raise4truept\hbox{$\sim$}}}$}\,}
\def\micron{\mbox{$\mu$m}}
\def \SAIT #1 #2 {{\em Mem.\ Soc.\ Astron.\ It.\/} {\bf #1}, #2}
\def \MESS #1 #2 {{\em The Messenger\/} {\bf #1}, #2}
\def \ASTRNACH #1 #2 {{\em Astron. Nach.\/} {\bf #1}, #2}
\def \AAP #1 #2 {{\em Astron. Astrophys.\/} {\bf #1}, #2}
\def \AAL #1 #2 {{\em Astron. Astrophys. Lett.\/} {\bf #1}, L#2}
\def \AAR #1 #2 {{\em Astron. Astrophys. Rev.\/} {\bf #1}, #2}
\def \AAS #1 #2 {{\em Astron. Astrophys. Suppl. Ser.\/} {\bf #1}, #2}
\def \AJ #1 #2 {{\em Astron. J.\/} {\bf #1}, #2}
\def \ANNREV #1 #2 {{\em Ann. Rev. Astron. Astrophys.\/} {\bf #1},
#2}
\def \APJ #1 #2 {{\em Astrophys. J.\/} {\bf #1}, #2}
\def \APJL #1 #2 {{\em Astrophys. J. Lett.\/} {\bf #1}, L#2}
\def \APJS #1 #2 {{\em Astrophys. J. Suppl.\/} {\bf #1}, #2}
\def \APSS #1 #2 {{\em Astrophys. Space Sci.\/} {\bf #1}, #2}
\def \ASR #1 #2 {{\em Adv. Space Res.\/} {\bf #1}, #2}
\def \BAIC #1 #2 {{\em Bull. Astron. Inst. Czechosl.\/} {\bf #1}, #2}
\def \JSQRT #1 #2 {{\em J. Quant. Spectrosc. Radiat. Transfer\/} {\bf
#1}, #2}
\def \MN #1 #2 {{\em Mon. Not. R. Astr. Soc.\/} {\bf #1}, #2}
\def \MEM #1 #2 {{\em Mem. R. Astr. Soc.\/} {\bf #1}, #2}
\def \PLR #1 #2 {{\em Phys. Lett. Rev.\/} {\bf #1}, #2}
\def \PASJ #1 #2 {{\em Publ. Astron. Soc. Japan\/} {\bf #1}, #2}
\def \PASP #1 #2 {{\em Publ. Astr. Soc. Pacific\/} {\bf #1}, #2}
\def \NAT #1 #2 {{\em Nature\/} {\bf #1}, #2}
\def \ACTA #1 #2 {{\em Acta Astron.\/} {\bf #1}, #2}
\documentstyle[]{l-aa}
\input psfig.tex

	\def\smallskip{\vskip 6pt}
	
	\def\M12{${\rm M_{12}}$}
\begin{document}

	\thesaurus{ }
\title{Modelling Intermediate Age and Old Stellar Populations in the
Infrared}

	\author{ A. Bressan $^1$, G. L. Granato $^1$, L. Silva $^2$}
	\institute{
		   $^1$ Astronomical Observatory, Vicolo dell'Osservatorio 5,
		   35122 Padova, Italy \\
		   $^2$ ISAS, Trieste, Via Beirut n.2-4,  34013 Trieste,
		   Italy
     }

	\offprints{A. Bressan }

	\date{Received July 1997; accepted September 1997}

	\maketitle
	\markboth{Isochrones in the infrared}{}

%%%%%%%%%%%%%%%%%%%%%%%%%%%%%%%%%%%%%%%%%%%%%%
%%%%%%%%%%%%%%%%%%%%%%%%%

\begin{abstract}
In this paper we have investigated the spectro-photometric properties
of the Asymptotic Giant Branch (AGB) stars and their contribution to
the integrated infrared emission in simple stellar populations (SSP).
Adopting analytical relations describing the evolution of these stars
in the HR diagram and empirical relations for the mass-loss rate and
the wind terminal velocity, we were able to model the effects of the
dusty envelope around these stars, with a minimal number of
parameters. After deriving simple scaling relations which allow us to
account for the metallicity of the star, we computed isochrones at
different age and initial metal content. We compare our models with
existing infrared colors of M giants and Mira stars and with IRAS PSC
data. The former data are fairly well reproduced by our models,
though a possible inadequacy of the adopted atmospheric models is
indicated. Though being characterized by different metallicity, the
isochrones follow a single path in the IRAS two-color diagram, fixed
by the composition and optical properties of the dust mixture. The
bulk of the data in the latter diagram is delimited by the curves
corresponding to a mixture of silicate grains and one of carbonaceous
grains. We also discuss the effects of detached shells of matter but
we do not take into account this phenomenon in the present
computations. 

Contrary to previous models, in the new isochrones the mass-loss
rate, which establishes the duration of the AGB phase, also
determines the spectral properties of the stars. The contribution of
these stars to the integrated light of the population is thus
obtained in a consistent way. We find that the emission in the mid
infrared is about one order of magnitude larger when dust is taken
into account  in  an intermediate age population, irrespective of the
particular mixture adopted. The dependence of the integrated colors
on the metallicity and age is discussed, with particular emphasis on
the problem of age-metallicity degeneracy. We show that, contrary to
the case of optical or near infrared colors, the adoption of a
suitable pass-band in the mid infrared allows a fair separation of
the two effects. We suggest intermediate redshift elliptical galaxies
as possible targets of this method of solving the age-metallicity
dilemma.

The new SSP models constitute a first step  in a more extended study
aimed at modelling the spectral properties of the galaxies from the
ultraviolet to the far infrared.
 
\keywords{stellar evolution: AGB stars, dust emission, isochrones --
galaxies: stellar populations, spectral evolution -- infrared: stars,
integrated colors}

\end{abstract}

\section{Introduction}

Effects of dust in the envelopes of Mira and OH/IR stars are usually
neglected in the spectrophotometric synthesis of a composite
population, on the notion that the contribution of dust enshrouded
stars to the integrated bolometric light is negligible.  While this
can be justified in very old systems, it is not the case among
intermediate age population clusters, whose brightest tracers are
indeed the asymptotic giant branch stars. In these stars, the light
absorbed by the dust at optical wavelengths is reemitted in a broad
region from a few to a few hundred microns, where it overwhelms the
stellar component. In addition to the thermal emission from the dust
particles there are characteristic lines from molecules and often
maser emission. Dust and molecular emission from these stars have
been interpreted as a signature of an expanding circumstellar
envelope, where gas particles reach the escape velocity giving rise
to significant mass-loss (Salpeter 1974a,b, Goldreich \& Scoville
1976, Elitzur, Goldreich \& Scoville 1976). Mass loss measurements
made by different authors and different techniques all agree with the
notion that Mira and OH/IR stars constitute the final nuclear
evolutionary phase where low and intermediate mass stars lose the
whole envelope and turns toward the fate of a white dwarf. The
process responsible of such a huge mass-loss is still unknown but,
nowadays, there is a growing amount of evidence from both the
observational and the theoretical point of view, that large amplitude
pulsations coupled with radiation pressure on dust grains play a
major role in determining the observed high mass-loss rates. From the
observational side there exists a tight correlation between the
mass-loss rate and the period of pulsation, even if the role of the
amplitude of pulsations has not yet been exploited (see Habing 1996
for a thorough review on the subject). On the other hand hydrodynamic
models show that large amplitude pulsations may levitate matter out
to a radius where radiation pressure on dust accelerates the gas
beyond the escape velocity (Bowen \& Willson 1991). Luminosity
functions of AGB stars in well studied populous clusters and fields
of the Large Magellanic Cloud (LMC) indicate that the mass loss along
this phase is far larger than that predicted by the usual Reimers law
which, on the contrary, successfully describes the evolution along
the red giant branch in old systems. A super-wind phase (Fusi-Pecci
\& Renzini 1976) is often invoked to account for the paucity of
bright AGB stars. A mass-loss rate exponentially increasing with
time, naturally evolving into a super-wind phenomenon, is obtained
both by the hydrodynamic models of Bowen \& Willson (1991), due to
the growth rate of the density scale height at the sonic point in the
envelope, as well as by the semiempirical treatment of Vassiliadis
and Wood (1993-VW), as a consequence of the growth rate of the period
of pulsation as the star climbs along the AGB losing its mass.
Finally only mass loss rates that include a super-wind phase, like
those described above, can account for the relation between the
initial stellar and final white dwarf mass.(e.g. Weidemann 1987).

In spite of the tight link between the mass--loss rate and the 
infrared emission among these objects very few attempts have been
made in providing a coherent picture of the photometric evolution of
an AGB star toward its final fate. In particular existing isochrones
do not account for the effects of dust and molecules around AGB stars
and are inadequate to study the photometric properties of star
clusters and galaxies when and/or where these stars contribute a
significant fraction of the light. To cope with this difficulty we
constructed a set of theoretical models which account for the effects
of the circumstellar envelope and obtained a new set of theoretical
isochrones particularly suited for the analysis of the infrared and
mid infrared data.

We stress in advance that our viewpoint is different from that
adopted in several previous studies devoted to the subject. Usually a
physically sound model is constructed and fitted to the available
observations in order to obtain the characteristic parameters of the
model such as the dust mass-loss rate, the effective temperature, the
grain composition and so on. The gas mass--loss rate follows from the
assumption of a value of the dust--to--gas ratio, or alternatively
this latter quantity is obtained when the gas mass--loss rate is
directly estimated from other measurements. Models are thus applied
and tuned to observations of specific objects.  This ensemble of data
constitutes a statistical basis for the parameterization of the
mass--loss rate as a function of the period of the stars, a relation
that has recently been widely adopted in model computations.

Here we will proceed along the opposite direction. We will assume
that the mass--loss rate of AGB stars is known as a function of the
basic stellar parameters (mass, luminosity, radius) either from
hydrodynamic models (e.g. Bowen \& Willson 1991) or from empirical
relations (VW). For a given dust to gas ratio the mass-loss rate in
dust is derived, and the spectrophotometric properties of the
envelope are computed from the given stellar spectrum and from the
geometry and velocity of the matter flow. A series of photometric
envelope models is then computed for several different basic
parameters and adopted whenever necessary in the construction of the
isochrones, in order to obtain both the colors of the single stars
and the integrated spectra.

The paper is organized as follows. Section two describes the envelope
model. Adopting a spherically symmetric stationary flow of matter the
radiative transport equation is solved to derive the extinction and
the emission of the dust as a function of the wavelength. The
reliability of the model is tested by comparing it to data assembled
from the literature. A monoparametric class of models is constructed
as a function of the optical depth of the envelope $\tau$ at 1$\mu$m
(thereinafter $\tau_{1}$) and a relation between the optical depth
$\tau_{1}$ and the basic stellar quantities (mass-loss rate,
expansion velocity
and luminosity) is obtained, in order to allow
the interpolation between two different stellar envelopes.

Section three describes the construction of the isochrones in the
theoretical HR diagram. Relations providing the mass-loss rate and
the 
expansion velocity are adapted from
Vassiliadis \& Wood (1993) with some minor modifications suggested by
the comparison of our envelope model with existing observations of
stars in the Magellanic Clouds and in our galaxy: in both the
super-wind mass-loss rate and the velocity period relation we include
a suitable dependence on the metallicity. 

In section four we describe the spectrophotometric properties of the
isochrones. For the evolutionary phases before the AGB, the method is
identical to the one adopted by Bressan et al. (1994). Along the AGB
the suitable envelope model is applied to the otherwise unaffected
stellar spectrum and absorption and reemission by dust is then
included in the model as it moves along the AGB phase. In this way we
are able to obtain consistent isochrones and corresponding integrated
spectra in the mid infrared dominion. We compare our isochrones of
different ages and metallicity with the IRAS two colors diagrams and
discuss the reliability of our sequence of envelope models. We
briefly discuss the difficulty of a monoparametric sequence of
envelopes with varying optical depth $\tau_1$ to interpret the IRAS
two--colors diagrams, already encountered by other authors (Bedijn
1987, Ivezic \& Elitzur 1995). The comparison with near infrared
colors of a sample of M giants and Miras shows that the inclusion of
the dust constitutes a significant improvement with respect to
previous models.

Section five is devoted to the spectro-photometric integrated
properties of the simple stellar populations (SSP). In particular we
stress the difficulty to disentangle age and metallicity effects from
the analysis of the optical and near IR colors alone. However we show
that extending the colors to the mid infrared may solve the
age-metallicity degeneracy at least in intermediate age systems.
Possible targets where the new models should exploit their highest
capabilities are thus intermediate redshift elliptical galaxies where
the stellar populations are only few Gyr old.

\section{The dust envelope model}

Radiative transfer in dusty shells around ``windy'' stars has been
previously considered by many authors (see Habing, 1996 and
references therein).  The most advanced approaches couple the
radiative transfer and the hydrodynamic equations of motion for the
two interacting fluids of the wind, the gas and the dust (Habing et
al. 1994; Ivezic \& Elitzur 1995).  This is necessary to achieve a
fully consistent solution, because the radiation pressure on dust
grains is widely believed to be the main driver of the high $\dot M$
observed in OH/IR stars.

However our interest here is to ``correct'' the stellar spectra
predicted by standard evolutionary tracks by including the effects of
dusty envelopes associated with AGB mass-loss.  We need therefore a
recipe, consistent with available observations, to associate the
fundamental parameters of the star, such as mass-loss,
escape-velocity, luminosity, radius and metallicity, with the
geometrical parameters of its dusty shell most relevant in
determining the spectral effects.
Along an isochrone the stellar parameters will change according to
some specified relations that will be discussed in the following
section, along with the procedure of assigning each point of the
isochrone the most adequate envelope model. Here we will assume that
the luminosity, the radius, the mass-loss rate, the escape velocity
and the dust to gas ratio of the model are known, and we will derive
the corresponding spectral properties of the dust envelope. 

We adopt spherical symmetry and, in a first order picture, an
outflow velocity $v_{\rm exp}$ independent of radius. Thus for a
given $\dot M$, the dust density outside the sublimation radius
$r_{in}$ wherein the grains are supposed to form suddenly, scales as
$r^{-2}$:
\begin{equation}
\rho_d(r)=\frac{\dot M \delta}{4 \pi v_{\rm exp}}\frac{1}{r^2}
\label{eq_rhorunsim}
\end{equation}
where $\delta$ is the assumed dust--to--gas mass ratio.
The envelope extends out to a radius $r_{out}=1000 \, r_{in}$.

We also test the effects of the presence of an expanding 0.02
M$_\odot$ shell of matter superimposed to the standard r$^{-2}$ law.
However these effects will not be taken into account in our SSP models 
and are only meant to illustrate uncertainties brought about by transient 
phenomena.
Adopting a gaussian profile for the density enhancement due to the
annulus alone, centered at $r_a<r_{out}$ (the outer radius of the
dust distribution) and with characteristic width $\Delta r_{a}$, the
total dust density becomes
\begin{equation}
\rho_d(r)=\frac{\dot M \delta}{4 \pi v_{\rm exp}}\frac{1}{r^2}
\left[1+f\times~e^{-\left(\frac{r-r_a}{\Delta r_{a}}\right)^2}
\right].
\label{eq_rhorun}
\end{equation}
It is straightforward to relate the constant $f$ to the mass
$\Delta M_{a}$ (=0.02 M$_\odot$) in the annulus:
\begin{equation}
\Delta M_a= \frac{\dot M \, f}{v_{\rm exp}} \int_{r_{in}}^{r_{out}}
e^{-\left(\frac{r-r_a}{\Delta r_{a}}\right)^2} dr
\label{eq_acc}
\end{equation}
To mimic an outward moving density enhancement, we computed three models
with $\Delta r_a=0.1 \, r_{out}$ and $r_a=0.2 \, r_{out}$, $r_a=0.5 \, r_{out}$ 
and $r_a=0.8 \, r_{out}$. 
Note that the corresponding density enhancement barely affects the
optical depth of the envelope, though it may affect the emission
in a given spectral region. Indeed
the fraction of column density due to the
enhancement in the annulus is $\simeq 3 \times 10^{-2} f$, $\simeq 7
\times 10^{-4} f$ and $\simeq 3 \times 10^{-4} f$, respectively, for
the three adopted values of $r_a/r_{out}$. On the other hand the
mass--loss rates relevant for this paper are $\gtrsim 10^{-5}
M_\odot/$yr while typical values of the outflow velocity are
$\leq 20$ km/s. Eq.\ \ref{eq_rhorun} yields then $f \leq 5$
which means that the optical depth is always dominated by the
standard $r^{-2}$ term. 
\begin{table}
\caption{Parameters of the adopted grain mixture.}
\centering
\begin{tabular}{lccccc}
$\phantom{1234567890123}$ & & & & \\
\multicolumn{1}{c}{Type} & $a\ [\micron]$ & $D\ [\mbox{gr/cm}^3]$ &
$X$ & 
$T_s$ [K]\\
\hline
\multicolumn{5}{c}{Mixture A} \\
Amorph.\ Sil.\ \dotfill &$ 0.1 $& 2.50 & 0.781 & 1000 \\
Silicate \dotfill &$ 0.03$& 2.50 & 0.189 & 1000 \\
Silicate \dotfill &$ 0.01$& 2.50 & 0.030 & 1000 \\
\hline
\multicolumn{5}{c}{Mixture B} \\
Amorph.\ Carb.\ \dotfill &$ 0.1 $& 2.26 & 0.398 & 1500 \\
Graphite \dotfill &$ 0.03$& 2.26 & 0.258 & 1500 \\
Graphite \dotfill &$ 0.01$& 2.26 & 0.344 & 1500 \\
\hline
\end{tabular}
\label{mix}
\end{table}

The parameters $\dot M$ and $v_{\rm exp}$ entering into Eq.\
\ref{eq_rhorun} are derived from the fundamental stellar parameters
as described in section 3. The remaining problem is to estimate the
dust to gas mass fraction $\delta$. It is widely believed that the
high mass loss rates in the AGB phase are driven by the transfer of
momentum of photons to the dust grains and then to the gas. Therefore
a close relationship between the dust abundance and the velocity of
the flow is expected. The problem has been studied carefully by
Habing et al.\ (1994), who confirmed that the terminal velocity of
the gas flow $v_{\rm exp}$ depends rather strongly on $\delta$. On
the other hand they found also that the flow reaches rather quickly a
velocity close to $v_{\rm exp}$, in keeping with our previous
assumption, and that the difference between the dust and the gas
velocity is positive and decreases with increasing $\dot M$. At the
high mass loss rates mostly relevant for this work the two velocities
are within
10 \%, and the results by Habing et al.\ can be well approximated by
the simple equation
\begin{equation}
\left( {v_{\rm exp}\over 21 \, \rm{km/s}} \right) \simeq
\left( {L\over 10^4 L_\odot} \right)^{0.35}
\left( {\delta\over 10^{-2} } \right)^{0.5}
\end{equation}
which can be inverted to yield
\begin{equation}
\delta \simeq 0.015 \, v_{\rm exp}^2 [{\rm km/s}] \,
\left( {L\over L_\odot}\right)^{-0.7}
\label{eq_delta}
\end{equation}

This set of equations gives the dust density at each position in the
dusty shell as a function of the stellar parameters.  Dust grains
absorb and scatter the stellar radiation, with an efficiency much
greater in the optical-UV regime than in the IR, and reradiate the
absorbed energy at IR wavelengths.  At the high mass loss rates
achieved in the super-wind phase, the involved densities imply that
the dust emission is self-absorbed, and therefore to compute the
emitted spectrum the radiative transfer problem must be solved.  The
numerical code we use is described in detail by Granato \& Danese
(1994) and we summarize here only its man features. The transfer of
radiation originating from a central source in an axisymmetric dust
distribution is solved with the lambda-iteration method: at each
iteration the local temperature of dust grains is computed from the
condition of thermal equilibrium with the radiation field estimated
at the previous iteration. The convergence is speeded up with respect
to this simple scheme following the prescriptions given by Collison
\& Fix (1991). Moreover, in the present case, taking advantage of the
spherical symmetry, the computing times are reduced by about two
orders of magnitude. The dust consists in a mixture of different
species of grains, which can be specified according to the needs. In
this paper, when dealing with OH/IR stars, we use the three silicate
grains of the six-grain (three silicates plus three carbonaceous)
model defined by Rowan-Robinson (1986) (mixture A). This model
provides a very simple but reasonably good description of the
absorption law of our Galaxy and of far-IR emission of galactic dust
clouds. As for carbon stars, we used the three carbonaceous grains of
the same model (mixture B). The characteristics of the dust grains,
type, dimension, density,  mass fraction and sublimation temperature
are displayed in Table \ \ref{mix}.
\begin{figure}
\psfig{file=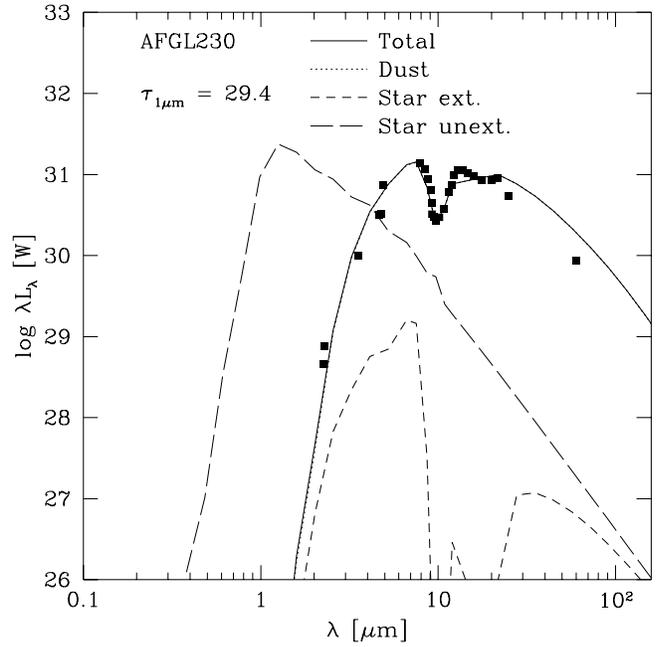,width=9cm}
 \caption{A fit to the star AFGL230 (data from Justtanont \& Tielens
 1992). The parameters of the fit are shown in Table \ref{fit}.
 Continuous line: emerging spectrum; dotted line, coinciding with the
 continuous line in this case: dust emission; long dashed line:
 original photospheric spectrum (T$_{eff}$=2500 K); short dashed
 line: extinguished photospheric spectrum.}
 \label{afgl30}
 \end{figure}
\begin{figure}
\psfig{file=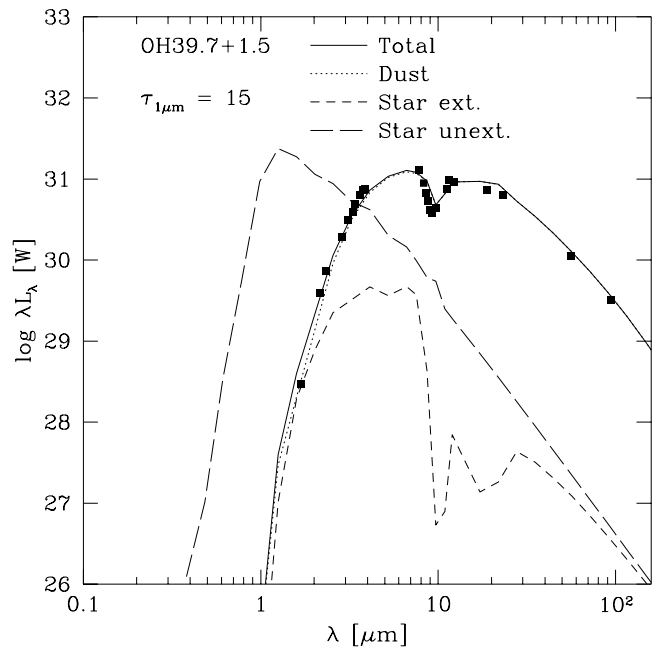,width=9cm}
 \caption{The same as in Fig.\ \ref{afgl30} but for OH~39.7+1.5 (data
 from Bedijn
 1987).}
 \label{oh39}
 \end{figure}
Our numerical code has been widely used in a different context,
namely models of dusty tori around AGNs (see Granato et al.\ 1997 and
references therein). Thus it is pretty well tested in the more
complex situation where spherical symmetry breaks down. As a further
check, we compared our results to the fits provided by other authors
to OH/IR stars spectra such as Justtanont \& Tielens (1992) and
Bedijn (1987). We found a quite good agreement in the fitting
parameters (Figs. \ref{afgl30}, \ref{oh39} and Table \ref{fit}
respectively) even if these authors have often introduced ad-hoc
modifications in the optical properties of dust, variable from object
to object, with the aim of giving a better description of the
strengths and shapes of the 10 and 18 micron silicate features. We
stress once again that our purpose is instead to have a good
description of the average properties of grains in circumstellar
shells.

% $\phantom{1234567890123}$ & & & & &\\
\begin{table}
\caption{Model parameters adopted to fit the energy distribution of
star AFGL230
and OH~39.7+1.5. The dust mass--loss rate $\dot M_d$ (M$_\odot$/yr)
has been
computed assuming
$v_{\rm exp}=15$ km/s; r$_{in}$ is in cm and $\rho_{d \ in}$, the
corresponding dust density
is in g/cm$^3$.
} 

\centering
\begin{tabular}{cccccc}
$\tau_1$ & Log(L/L$_\odot$) & Te & r$_{in}$ & $\rho_{d \ in}$ & $\dot
M_d$\\
\hline
\multicolumn{6}{c}{AFGL230} \\
29.4 & 4.715 & 2500 & 7.1E+14 & 5.9E-18 & 8.9E-07\\
\multicolumn{6}{c}{OH~39.7+1.5} \\
15.0 & 4.715 & 2500 & 7.1E+14 & 3.0E-18 & 4.5E-07\\
\hline
\end{tabular}
\label{fit}
\end{table}

We will now show that $\tau_1$ can be expressed as a function of
known parameters of the star. This result holds true also in 
presence of a density enhancement (eq. \ref{eq_rhorun}),
because the corresponding column density variation is
of the order of 10\% at most. On the contrary the spectral properties
of the envelope depend on both $\tau_1$ and f, but the latter effect
requires a detailed modelling that is beyond the scope of the present
work.
Thus, integrating for the sake
of simplicity Eq.\ \ref{eq_rhorunsim} from $r_{in}$ to $r_{out}$ we
have
\begin{equation}
\tau_1={\dot M \, \delta \, k_1 \over 4 \pi \, v_{\rm exp}} {1 \over
r_{in}}
\label{eq_tmptau}
\end{equation}
where the term $1/r_{out}$ has been neglected with respect to
$1/r_{in}$ and $k_1$ is the dust opacity at 1 $\mu$m. On the other
hand the dust sublimation radius is (Granato \& Danese 1994)
\begin{equation}
r_{in}=b \, L^{1/2}
\end{equation}
where the factor $b$ depends mainly on the dust composition (through
its melting temperature $\sim 1000$ K for silicates and $\sim 1500$ K
for graphite), the size distribution and only weakly on the shape of
the stellar spectrum. Typical values of $b$ are 2--3$\times$10$^{12}$
for
silicates mixture and 1--2$\times$10$^{12}$ for carbonaceous mixture.

Using this equation and Eq.\ \ref{eq_delta} into Eq.\ \ref{eq_tmptau}
we get
\begin{equation}
\tau_1= \alpha {\dot M \, v_{\rm exp} \over L^{1.2}}
\label{eq_tau}
\end{equation}
where $v_{exp}$ is in
km/s, $\dot M$ in M$_{\odot}$/yr and $L$ in L$_{\odot}$.
The  quantity $\alpha$, which incorporates the dependence upon  k$_1$
and b, is affected only weakly by the spectrum of the illuminating
star. For hot stars
emitting mainly in optical UV regime, where the optical properties
of dust do not show orders of magnitude variations with the
wavelength, the latter dependence would be quite negligible for
practical purposes. By converse in our case (cold AGB stars) the
situation is a bit less simple.  We checked however with the code
that adopting $\alpha=2.32 \times 10^9$ for the silicates mixture 
and $\alpha=9.85 \times 10^9$ for the carbonaceous mixture, the
errors in $\tau_1$ estimated from Eq.\ \ref{eq_tau} are kept within 5
\% for stars with $T_{\rm eff}$ in the range 2500-4000.
This is a fair approximation because models with $\tau_1$
differing by less than 5-10 \% produce almost identical spectra.

In conclusion, with the machinery described in this section, we 
computed a sequence of envelope models, providing 
both the extinction ($\tau_\lambda$) and emission ($D_\lambda$) 
figures, as a function of $\tau_{1}$ for two different dust mixtures. 

We also derived a relation between $\tau_{1}$ and the luminosity, the
expansion velocity $v_{\rm exp}$ and the mass loss rate $\dot M$ of the star.
How we obtain these quantities along an isochrone,  is described in the
next section.

\section{The isochrones}

To investigate the infrared properties of SSPs we coupled the dusty
envelope model described in the previous section with the large
library of stellar evolutionary tracks existing in Padova. These
tracks are computed for different values of the initial metallicity,
keeping constant and equal to the solar partition the relative
proportion of the metals. The initial mass of the evolutionary tracks
is in the range 0.6 to 120 M$_\odot$ corresponding to ages from few
million yr.\ to several billion yr. Since the main interest of this
paper is to focus on the photometric properties of AGB stars,
isochrones have been computed only for intermediate age and old
stellar populations, thus between say $ 10^8$ yr to $16\cdot 10^9$yr.
The initial chemical composition of the evolutionary sequences
adopted here  (Bressan at al. 1993 and Fagotto et al. 1994a, b), is
[Z=0.004, Y=0.24], [Z=0.008, Y=0.25], [Z=0.02, Y=0.28], and [Z=0.05,
Y=0.352], respectively.

The evolution of low
mass stars is computed at constant mass and mass loss is applied
during the procedure of isochrones construction. The mass-loss rate
along the RGB is parameterized by the usual Reimers formulation with
$\eta$=0.45.

Since the Padova models do not extend into the thermally pulsating
AGB phase (TP-AGB), an analytic description is adopted to complete
the isochrones up to the phase of the formation of the planetary
nebula. We adopt the analytic procedure of Bertelli et al. (1994)
which rests on some well defined relations provided by full numerical
calculations, namely the core-mass luminosity relation and the
reference AGB locus in the Hertzsprung-Russell (HR) diagram as a
function of the mass and metallicity of the star (see e.g. VW).

\subsection{Mass-loss rate along the AGB}

The mass-loss rate along the AGB is a key parameter for the evolution
of these stars because it affects the lifetime and the average
luminosity of the phase. Both these quantities bear on the
contribution of the whole phase to the integrated light because the
larger the mass-loss rate the shorter the lifetime of the phase, the
dimmer its brightest stars and finally the lower the contribution to
the total light of the population. The mass-loss rate also affects
the core mass of the subsequent post-AGB stars, hence the ultraviolet
properties of the stellar population.

A great effort has been recently devoted to clarify the role of the
mass-loss on the evolution of stars along the AGB. It has been shown
(Bowen \& Willson 1991, VW, Bl\"ocker 1995a, Groenewegen \& De Jong
1994, Marigo et al. 1996, 1997) that the mass-loss rate rises almost
exponentially with time during the AGB phase and that sooner or later
it turns into a superwind that completely evaporates the envelope of
the star, leaving a bare core which then evolves toward very high
temperatures. All bright AGB stars are known to be variables with
amplitudes that may reach about a couple of magnitudes, and infrared
observations of Mira and OH/IR stars show the characteristic emission
features of a dusty envelope suggesting that pulsations and dust may
play an important role in the mass-loss process.  However the actual
mechanism which drives the mass-loss is still presently not well
understood. Moreover mass-loss rate determinations, derived mainly
from IR and millimetric measurements, are characterized by a
significant uncertainty, related to the unknown distance of the
source, the assumed dust to gas ratio and the expansion velocity.

Recent hydrodynamic calculations (Bowen \& Willson 1991, Willson et
al. 1995) show that the shock waves generated by large amplitude
pulsations of AGB stars levitate matter out to a radius where dust
grains can condensate; from this point, radiation pressure on grains
and subsequent energy redistribution by collisions, accelerate the
matter beyond the escape velocity.  In this model, the exponential
growth of the mass-loss rate is entirely due to the variation of the
density scale height below the acceleration region (Bowen \& Willson
1991) as the star climbs along the AGB. On the other hand a simple
fit to the relation between the mass--loss--rate and the period along
with the analytical relations between period, mass, luminosity and
effective temperature of an AGB star (see e.g. VW) show that the
mass-loss rate increases exponentially with the luminosity and, above
a critical threshold, turns into a superwind limited only by the
reservoir of the momentum in the radiation field.

Here we adopt the formalism of VW. Following their empirical
relation, the mass loss rate $\dot M$ grows exponentially with the
pulsation period $P$ until a constant upper limit is reached at a
period of about 500 days, which corresponds to the superwind phase.
The relation between the mass-loss rate and the period has been
derived from observational determinations of mass-loss rates for Mira
variables and pulsating OH/IR stars both in the Galaxy and in the
LMC. The two regimes are equivalent for a period of about 500 days
for the solar composition.

The adopted relations are:
% %\log \dot M = -11.4 + 0.0125(P - 100*(M - 2.5))
\begin{equation}
\log \dot M = -11.4 + 0.0123 P
\label{mlr1}
\end{equation} %

Here and in the following, $\dot M$ is given in units of $M_{\odot}$
yr$^{-1}$, the stellar luminosity $L$ is expressed in $L_{\odot}$,
the pulsation period $P$ in days, $c$ is the speed of light (in km
s$^{-1}$) and $v_{\rm exp}$ (in km s$^{-1}$) is the terminal velocity
of stellar wind.

VW express the mass-loss rate in the superwind phase by equating the
final mass momentum flux $\dot{M} v_{\rm exp}$ to the momentum flux
of the entire stellar luminosity, according to the
radiation-driven-wind theory (Castor et al. 1975):

%\dot M = 8.088 \:\:10^{-9}\frac{Z}{0.008}\times\frac{L}{v_{\rm exp}}
%
\begin{equation}
\dot M = 6.07023 \:\:10^{-3}\beta\frac{L}{c v_{\rm exp}}
\label{mlr2}
\end{equation}

adopting $\beta$ = 1. While the Galactic OH/IR stars suggest a value
of the order unity (Wood et al. 1992), there is no reason why the
same value should hold true for other environments such as the LMC
because, as clarified by Netzer and Elitzur (1993), the momentum
transfer cannot constrain the value of $\beta$. Furthermore Wood et
al. (1992) found that the expansion velocity is clearly lower in more
metal poor stars and concluded that by adopting a constant value for
$\beta$ one would predict that the mass-loss increases at decreasing
metallicity, contrary to what is observed.

We thus included a metallicity dependence of the mass-loss rate in
the superwind phase by imposing that $\beta$ scales linearly with the
metallicity Z and calibrated this dependence by fitting the 
infrared spectrum of the OH/IR star TRM 60 in the LMC( Groenewegen et
al. 1995). With our envelope model we estimate a value of $\tau_1$ of
8.  Assuming a distance modulus for LMC (m-M) $_0$=18.5, the
luminosity of the star is $5.4 \; 10^4 \, {\rm L}_{\odot}$. Since the
observed expansion velocity of TRM 60 is 12 km s$^{-1}$ (Wood et al.
1992) we get a dust mass-loss rate of $1.1 \; 10^{-7} {\rm
M}_{\odot}/{\rm yr}$. To convert it in gas mass-loss rate we derive
from our scaling law (Eq.\ \ref{eq_delta}) a dust to gas ratio of
$10^{-3}$ so that, assuming that this object is in a superwind phase,
we get $\beta=1.13$. If we further adopt a metallicity of Z=0.008 for
this star (an average value for the LMC) we have:

\begin{equation}
\dot M = 6.07023 \:\:10^{-3}\frac{L}{c v_{\rm
exp}}\times{1.13}\frac{Z}{0.008}
\label{mlr3}
\end{equation}
For the Galactic OH/IR stars this relation implies that $\beta\sim$2,
which is
a suitable value (Wood et al. 1992).

As far as the expansion velocity is concerned, VW provide the
following relation between the velocity $v_{\rm exp}$, (in km
s$^{-1}$) and the period of the star
\begin{equation}
v_{\rm exp} = -13.5 + 0.056 P
\label{vp1}
\end{equation}
with the additional constraint that $v_{\rm exp}$ is greater than $3$
km s$^{-1}$, a lower limit typical of Mira and OH/IR stars.
Furthermore VW constrain the velocity to be less than the average
value of $15$ km s$^{-1}$, again a typical value for such stars.
However, as shown by Wood et al. (1992), there seems to be a trend
between the expansion velocity and the period even for P $\geq$ 500
days (the period corresponding to $v_{\rm exp}$=15 km s$^{-1}$ in the
above equation) and, perhaps more important, there seems to be a
trend with the metallicity of the star, because the observations
suggest that $v_{\rm exp}(LMC)$=0.5-0.6~$\times$ $v_{\rm
exp}(Galaxy)$. We thus let the maximum velocity to depend on the
period and the metallicity of the star according to the relation
\begin{equation}
v_{\rm exp} \leq {6.5}\frac{Z}{0.008} + 0.00226 P
\label{vp2}
\end{equation}
The slope of this relation has been derived from a linear fit to the
data for the Galactic OH/IR stars within 1$^o$ of the Galactic plane
(Fig. 6 of Wood et al. 1992), while the metallicity dependence of the
zero point rests on the assumption that the few data for the LMC
OH/IR stars in the same figure obey the same relationship, however
scaled to a lower metallicity Z=0.008 instead of Z=0.02.

Finally the pulsation period $P$ is derived from the
period-mass-radius relation (see Eq. 4 in VW) which is obtained by
assuming that variable AGB stars are pulsating in the fundamental
mode: %
\begin{equation}
\log P = -2.07 + 1.94\:\: \log R - 0.9\:\: \log M
\label{pmlt}
\end{equation}
where the period $P$ is given in days and the stellar radius $R$ and
mass $M$ are expressed in solar units.

A final comment concerns the validity of the core-mass luminosity
relation, which has been assumed here as one of the relations adopted
to compute the analytic TP-AGB phase. It is well known that hot
bottom burning in stars of initial mass larger than about 4~M$_\odot$
causes a brightening of about 20\% with respect to the usual
core-mass luminosity relation. However the above effect disappears
when the star reaches the superwind phase (see e.g. VW) which is the
most relevant for the present purpose. We also neglected the
modulation of the luminosity introduced by thermal pulses (e.g.
Marigo,
Bressan \& Chiosi 1996, 1997) on the notion that also this effect
constitutes a higher degree of approximation which is beyond the
current scope as we need to keep only a minimum number of free
parameters at work.

\subsection{ The PAGB stars} As a consequence of the superwind phase
the star eventually reaches a maximum luminosity after which it loses
most of its envelope and rapidly evolves toward the so called
Post-AGB path (PAGB). Thereafter the star reaches a maximum
temperature at almost constant luminosity and then it cools and dims
along the White Dwarf cooling sequence. The observed initial-final
mass relation of PAGB stars (Weidemann 1987) constitutes a further
test for the AGB phase and supports the kind of mass-loss rate
adopted here (VW, Bertelli et al. 1994)

PAGB stars have been suggested as one of the main contributors to the
integrated ultraviolet light of nearby elliptical galaxies. In the
bulge of M31 their estimated contribution to UV light amounts to
$\simeq$ 20\%. However in M32 their number per unit UV light, as seen
from the same HST FOC 1550 optical combination, is an order of
magnitude lower. This indicates that age and metallicity effects may
significantly affect the contributions of the PAGB stars that are
ultimately related to the AGB evolution.

Characteristic evolutionary paths of PAGB stars of different mass and
composition have been computed by Paczy\'nski (1971), Sch\"onberner
(1981), Iben (1984), Bl\"ocker \& Sch\"onberner (1991), Fagotto et
al.\ (1994) and Bl\"ocker (1995b). For the present isochrones we
assembled the data from Sch\"onberner (1981), Bl\"ocker \&
Sch\"onberner (1991), Fagotto et al.\ (1994), and Bl\"ocker (1995b).

\section{Isochrones in the infrared}
\begin{figure*}
\centering
\psfig{file=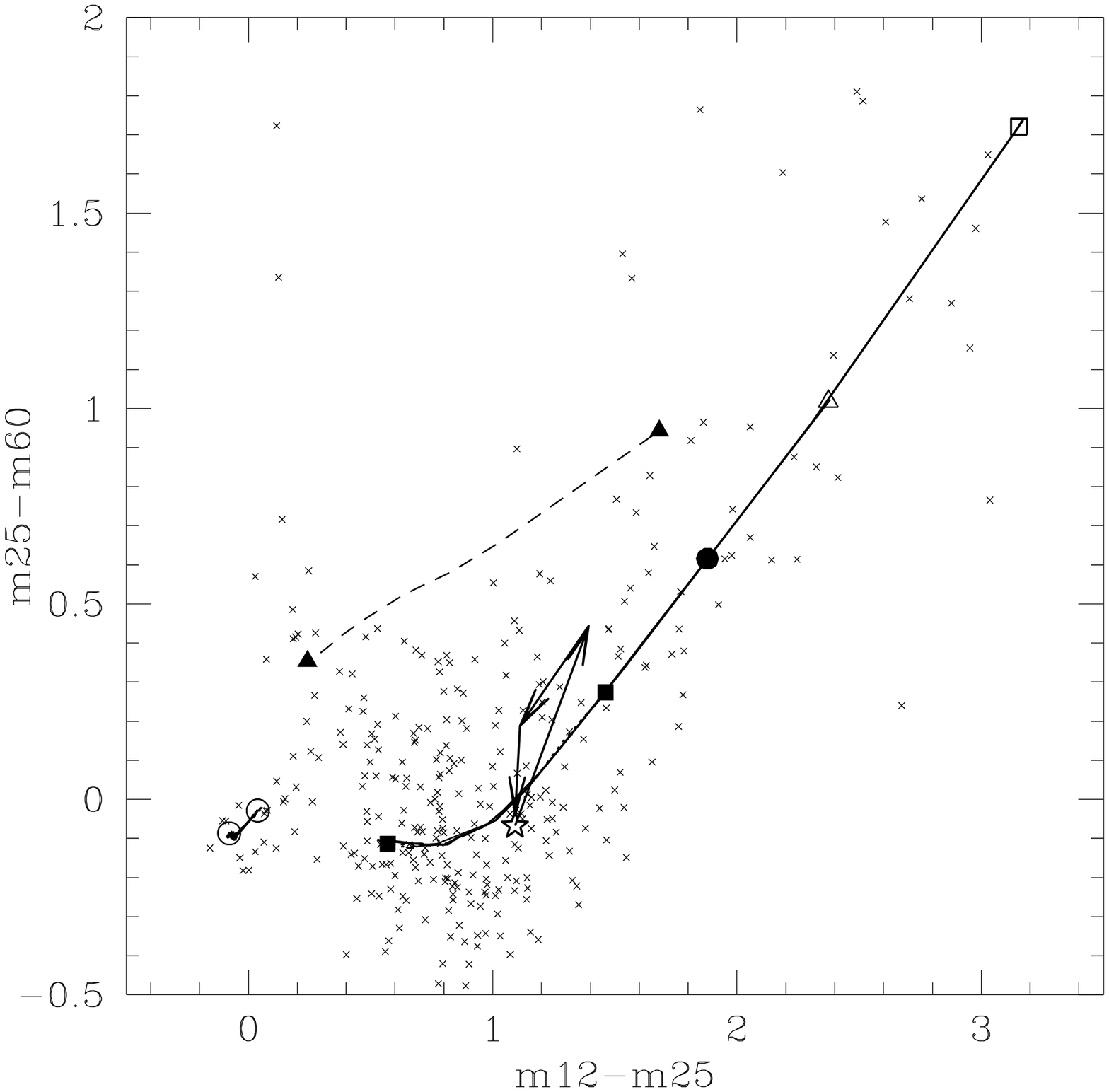,width=15cm}
\caption{Two colors diagram of the sample of IRAS sources defined by
Van der Veen \& Habing (1988). Following Ivezic and Elitzur (1995)
the sources with cirr3/F(60)$>$2 have been excluded in order to
minimize contamination by cirrus emission. Superimposed are 1.5 Gyr
isochrones of different metallicity and a model of an AGB star
with T$_{eff}$=2500K, LogL/L$_\odot$=4, $\tau_1$=1 and 
$v_{\rm exp}$=15 km/s (empty star), surrounded by an expanding shell
of 0.02M$_\odot$ at 0.2, 0.5 and 0.8 r$_{out}$ (arrows). See
text for more details.}
\label{iras2}
\end{figure*}

Transformations from the theoretical to the observational plane were
performed by making use of the stellar spectral library of Bressan et
al.  (1994), Silva (1995) and Tantalo et al. (1996). The core of the
spectral library is the atlas by R. Kurucz (1992). At temperatures
higher than ${\rm T_{eff}} = 50,000$ K pure black-body spectra are
adopted, whereas for stars cooler than ${\rm T_{eff}} = 3500$ K the
catalog of stellar fluxes by Fluks et al. (1994) is implemented. The
latter library is based on stars of solar metallicity (${\rm Z \sim
0.02}$) whereas the library of SSPs presented here spans the range of
metallicity ${\rm 0.004 \leq Z \leq 0.05 }$. To account for a
dependence on the metal content even for M giants, we adopted the
same library at different metallicities but assigned the spectral
class, identified by the (V-K) color, adopting the (V-K)-${\rm
T_{eff}}$ relation of Bessell et al. (1991) which depends on the
metallicity.

To account for the effect of dust along the TP-AGB we proceed in the
following way. For a given temperature, luminosity, mass and
metallicity, making use of Eqs.\ \ref{eq_tau} (and \ \ref{eq_acc},
when required), we derive the value of $\tau_{1}$ and interpolate in
our sequence of envelope models, obtaining both the extinction
($\tau_\lambda$) and emission ($D_\lambda$) figures. The
characteristic stellar spectrum is then modified according to the
relation
\begin{equation}
F_\lambda = F_\lambda \times e^{-\tau_\lambda} + {D_\lambda}
\times L
\label{ftd}
\end{equation}
where the first factor in the RHS represents the extinction while the
second one represents the dust emission.
A final renormalization of the whole
spectrum to the $\sigma{T^4}$ relation is performed but it usually
amounts to a negligible correction, indicating that the whole
procedure is sufficiently accurate.

Before proceeding further we compare our results with the infrared
colors of Mira and OH/IR stars and discuss the reliability of our
selected circumstellar envelope models. Fig. \ref{iras2} shows
several isochrones of 1.5 Gyr and different metal content in the
two--colors IRAS diagram, superimposed to the sample of IRAS sources
defined by Van der Veen \& Habing (1988).
The original sample includes about 1400 
sources for which $[60]-[25]<0$ and $[25]-[12]<0.6$, these limits
define the regions occupied by galactic late type stars in the IRAS
two color diagram.  In the above notation
[$\lambda_2$]-[$\lambda_1$]=log(F$_{\lambda_2}$/F$_{\lambda_1}$).
However,  in order to minimize contamination by cirrus emission, we
excluded all the sources with cirr3/F(60)$>$2 (Ivezic and
Elitzur, 1995). The sample plotted in Fig. \ref{iras2} reduces to
about 300 sources.

A thorough discussion of this diagram together with the regions
occupied by galactic Mira and OH/IR stars can be found in Habing
(1996). Fig. \ref{iras2} reports data and models in form of
magnitudes. Theoretical IRAS monochromatic fluxes were obtained by
convolving the stellar spectral energy distributions with the proper
transmission curve as detailed in Bedijn (1987).  IRAS
magnitudes were thus derived according to 
\begin{equation} M_i =
-2.5\times Log(S_i)+2.5\times Log(S_{0_i}) 
\label{miras}
\end{equation} 
where S$_i$ is the flux in Jansky and the constants
$S_{0_i}$ are derived from the IRAS PSC-explanatory supplement (1988)
and are 28.3\ Jy, 6.73\ Jy, 1.19\ Jy and 0.43 for the 12, 25, 60 and
100 {$\mu$}m passband respectively. This procedure is adopted to
facilitate the construction of mixed optical-IR two color diagrams as
described below.

Turning now to Fig. \ref{iras2} we notice that, before reaching the
AGB phase, the isochrones are confined in a small region delimited by
two open circles, with  both (m12-m25) and (m25-m60) $\simeq$0.
These figures correspond to the Rayleigh-Jeans spectral regime and
indicate that this approximation is fairly accurate till the most
advanced phases of the isochrone, where dust comes into play. As the
star reaches the thermally pulsating AGB phase the effect of
mass-loss suddenly increases and the circumstellar envelope starts to
modify the spectral emission. The first envelope model corresponds to
$\tau_{1}$=0.01 and it is indicated by a filled square at (m12-m25)
$\simeq$0.6 and (m25-m60) $\simeq$-0.1 for the case of the silicate
mixture and by a filled triangle at (m12-m25) $\simeq$0.2 and
(m25-m60) $\simeq$0.35 for the carbonaceous mixture, respectively.

Fig. \ref{iras2} depicts isochrones of different metallicity but they
are barely distinguishable in this diagram. This is because our
envelope models are characterized by a single parameter, the optical
depth, so that different conditions resulting in the same value of
$\tau_1$, such as metallicity, mass-loss rate and expansion velocity,
provide the same dust envelope and in turn give rise to the same
overall spectrum. In contrast, a noticeable feature of this diagram
is the dispersion of the data. Ivezic \& Elitzur (1995) thoroughly
discussed this problem and they were able
to show that the bulk of the data are delimited by the $\tau_1$
curves corresponding to grains of either pure silicates or graphite.
The thick dashed line in Fig. \ref{iras2} refers to an isochrone we
computed accounting for carbonaceous grains only, characteristic of C
rich stars, and it is meant to illustrate the effects of the
variation of the chemical composition of the dust. Thus we confirm
the results of Ivezic \& Elitzur (1995). 
Fig. \ref{iras2} also depicts the path an AGB star would follow
during a shell ejection of 0.02\, M$_\odot$, and it is meant to
illustrate additional effects possibly at work which 
however, are difficult to predict on the basis of simple stellar
parameters.

It is remarkable that 
isochrones of different metallicity and age reach different maximum
values of $\tau_1$. 
The larger the metallicity the higher the value of $\tau_1$ and the
further the isochrone extends in the IRAS two color diagram. The
largest value of $\tau_1$ reached by models of different metallicity
are indicated in Fig. \ref{iras2} by a filled circle ($\tau_1$=4.4),
a filled square ($\tau_1$=11.1), an open triangle ($\tau_1$=22.3) and
an open square ($\tau_1$=57.5) for the metallicity Z=0.004, Z=0.008,
Z=0.02, Z=0.05 respectively. The filled triangle at (m12-m25)
$\simeq$1.7 and (m25-m60) $\simeq$0.9 refers to the case of Z=0.02
and a mixture of carbonaceous dust grains ($\tau_1$=99).

In summary, while we analyzed both the case of a different mixture
and a shell enhancement, we consider the ability of our simple models
to reproduce the bulk of the data in the IRAS two colors diagram as a
meaningful test of reliability, also in view of the errors on the
quoted fluxes ($\simeq 5\%$) reported in the IRAS PSC-explanatory
supplement (1988).
\begin{figure}
 \psfig{file=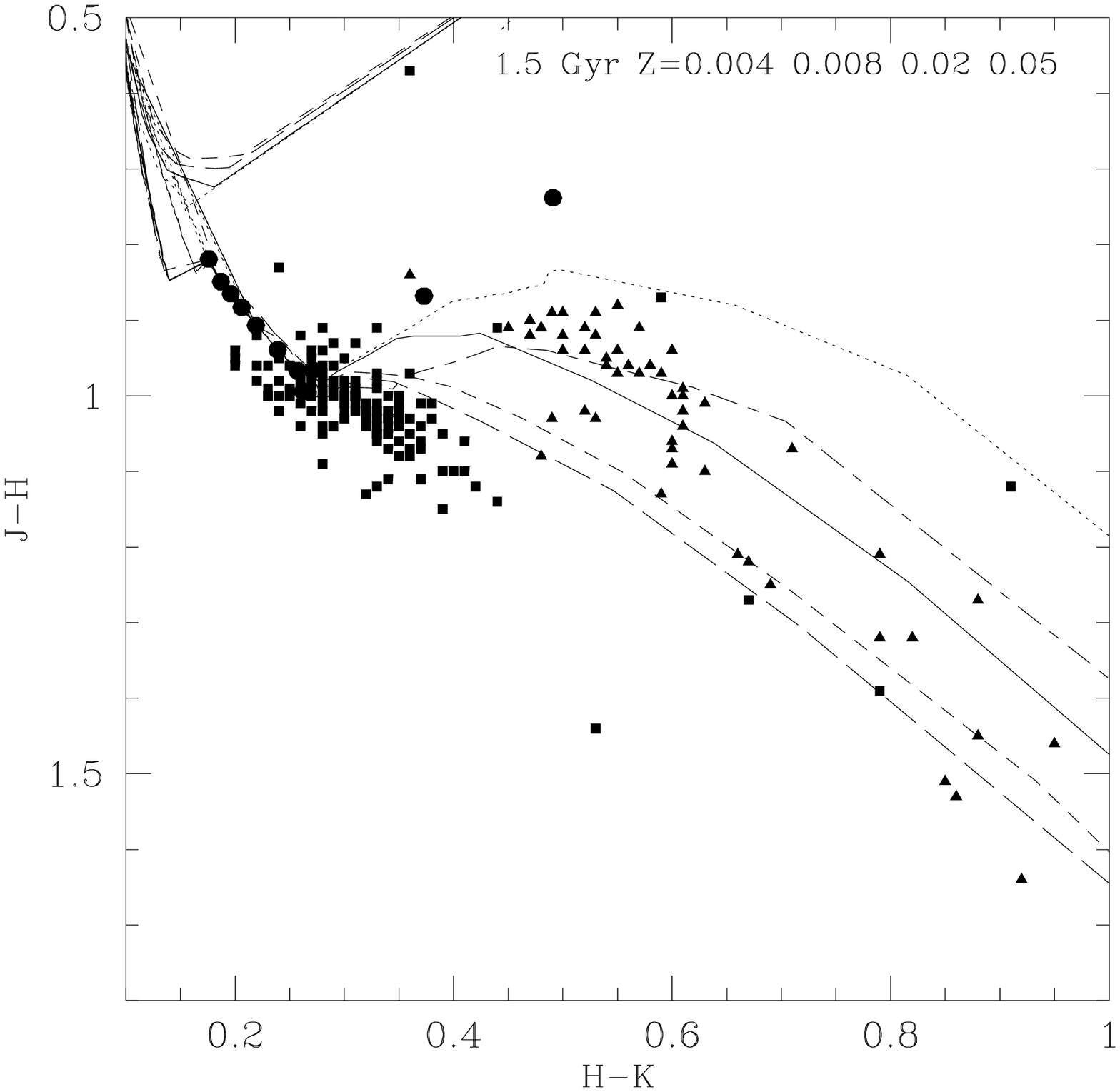,width=9.truecm}
 \caption{(J-H)-(H-K) two color diagram of a sample of galactic M
 (squares) and Mira (triangles) stars from Whitelock et al. (1994,
 1995). Superimposed are some selected isochrones of 1.5 Gyr with
 metallicity Z=0.004 (long-dashed line), Z=0.008 (short-dashed line),
 Z=0.02 (continuous line) and Z=0.05 (dotted line) for the mixture A,
 and Z=0.02 (long-short dashed line) for the mixture B. Large filled
 dots indicate the location of the adopted M giant atmospheric
 models.}
 \label{mirajhk}
 \end{figure}
\begin{figure}
 \psfig{file=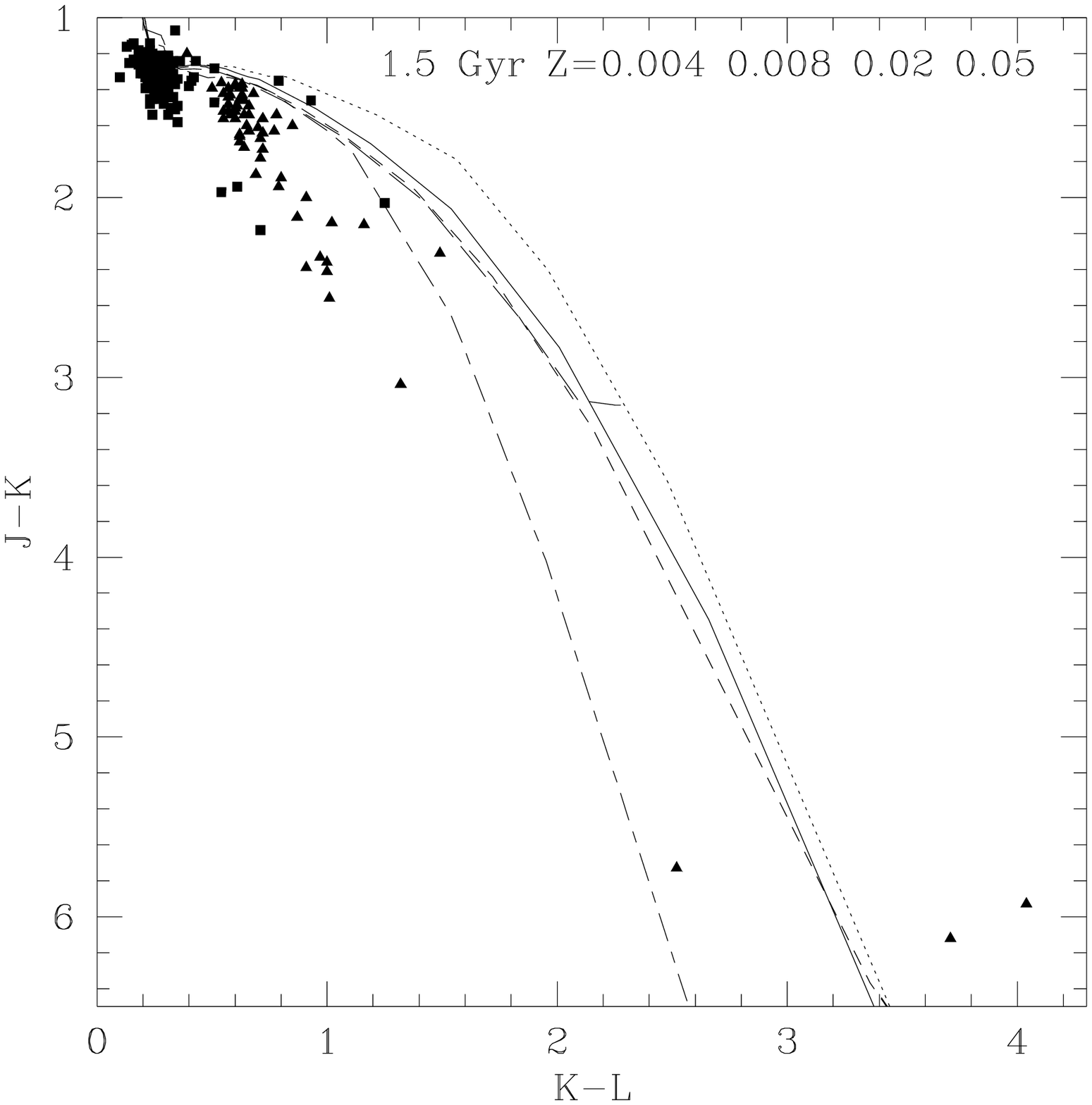,width=9.truecm}
 \caption{Same as in Fig \ref{mirajhk} but for the (J-K) and (K-L)
 colors} \label{mirajkl}
 \end{figure}

To check the behavior of our models in the near infrared region of
the spectrum, we have also compared our isochrones with a sample of M
giants (filled squares) and Miras (filled triangles) in the Southern
Polar Cap, for which Whitelock et al. (1994, 1995) obtained J, H, K
and L magnitudes. Fig. \ref{mirajhk} shows the two colors
(J-H)--(H-K) diagram of M giants and Miras in the Southern Polar Cap
(Whitelock et al. 1994, 1995) superimposed to some selected
isochrones at different age and metallicity. Both the bulk of the M
giants at (J-H) $\simeq$1 and (H-K)$\simeq$0.3 and those of Miras
sample are well reproduced by the models. The fit to the latter
sample however depends on the adopted temperature scale, in
particular to the temperature assigned to the more advanced M
spectral types of the Fluks atlas. Only by reaching the most advanced
spectral types during the superwind phase we are able to match the
Miras in Fig.  \ref{mirajhk}, as can be seen from the location of our
adopted M giant spectra (large filled circles in the figure).
Assigning a metallicity on the basis of the isochrone fitting of near
infrared colors is thus a quite delicate process, whose reliability
must be clearly improved by more accurate atmospheric models.  Note
that in our case the bulk of Mira stars are well fitted by models of
solar metallicity, while it is impossible to assign a metallicity to
the M giants.  Finally notice that the highly reddened data are
relatively well fitted by our dusty models. On the contrary
isochrones based on pure photospheric models would never be able to
fit the data, as can be seen from the location of late type giants of
our spectral library.

Fig. \ref{mirajkl} is analogous to Fig \ref{mirajhk} but for the
colors (J-K)--(K-L). Here the fit to the Mira sample is poor,
suggesting that the problem for the latter stars may be related to
the adopted atmospheric models (see also Bessell et al. 1991 and
Whitelock et al. 1994). However, as in the previous case,  the fit
improves significantly at high optical depth.
\begin{figure}
\psfig{file=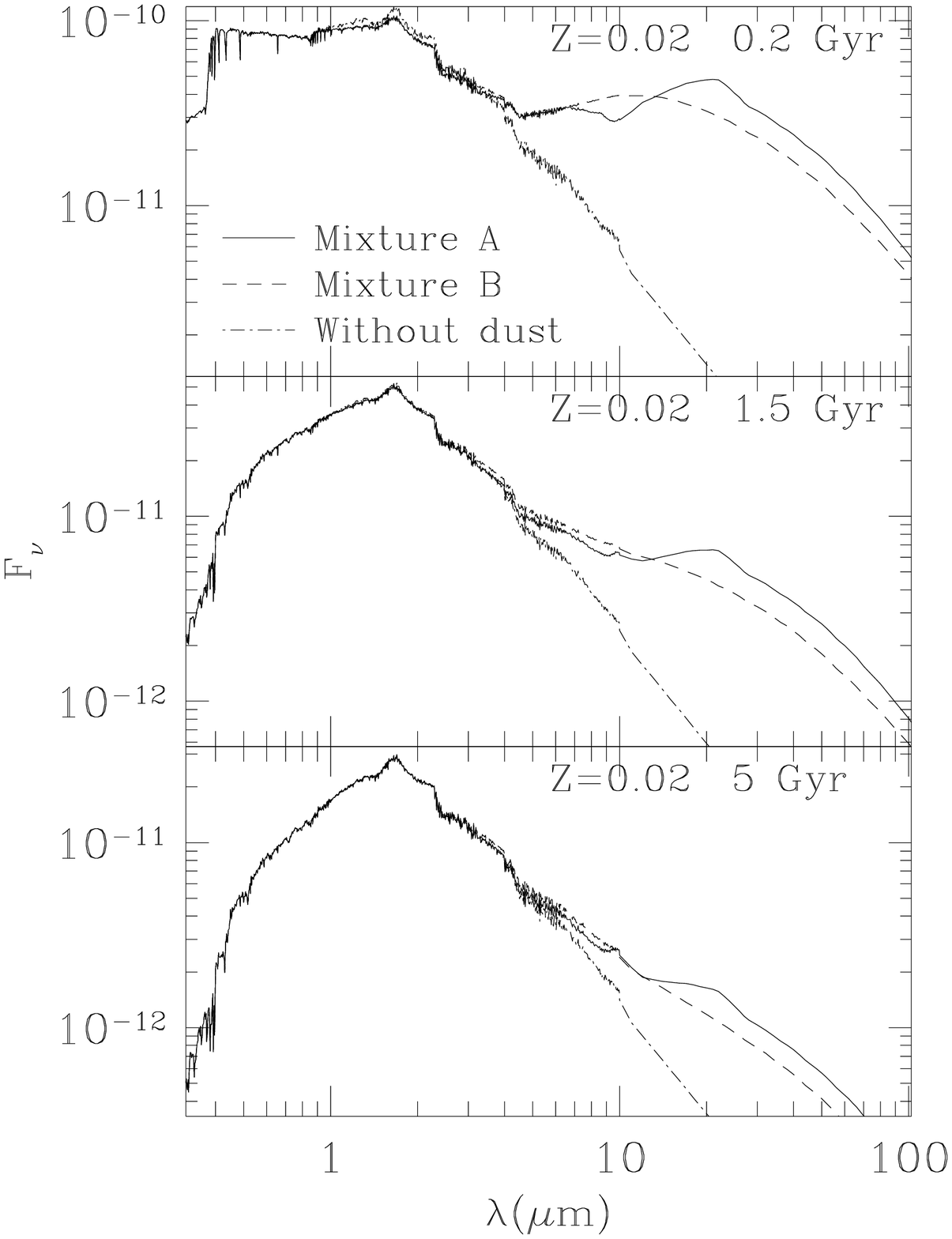,height=12truecm}
\caption{Integrated spectrum F$_\nu$ vs. $\lambda$ in $\mu$m of 
 SSPs of 0.2 Gyr 1.5 Gyr and 5 Gyr and solar composition. Three cases
 are depicted. In the first case
 dust is not taken in to account (dot-dashed line), in the second
 dust is a mixture of silicates grains (continuous line), and in the
 third dust is a mixture of carbonaceous grains (dashed line).}
 \label{ssp02g}
 \end{figure}

\section{Integrated  colors: the age-metallicity degeneracy}

Once the spectrum of a single star is obtained it is weighted by the
appropriate number of stars in the elemental interval of the
isochrone and summed up to obtain the integrated spectrum of the SSP.
Throughout this paper we adopted a Salpeter initial mass function
\begin{equation}
\phi(M) dM \propto M^{-x} dM
\label{imf}
\end{equation}
with x=2.35. The integration is performed from M$_{inf}$=0.15 up to
the largest initial mass which still contributes to the integrated
spectrum of the isochrone.
\begin{figure}
\psfig{file=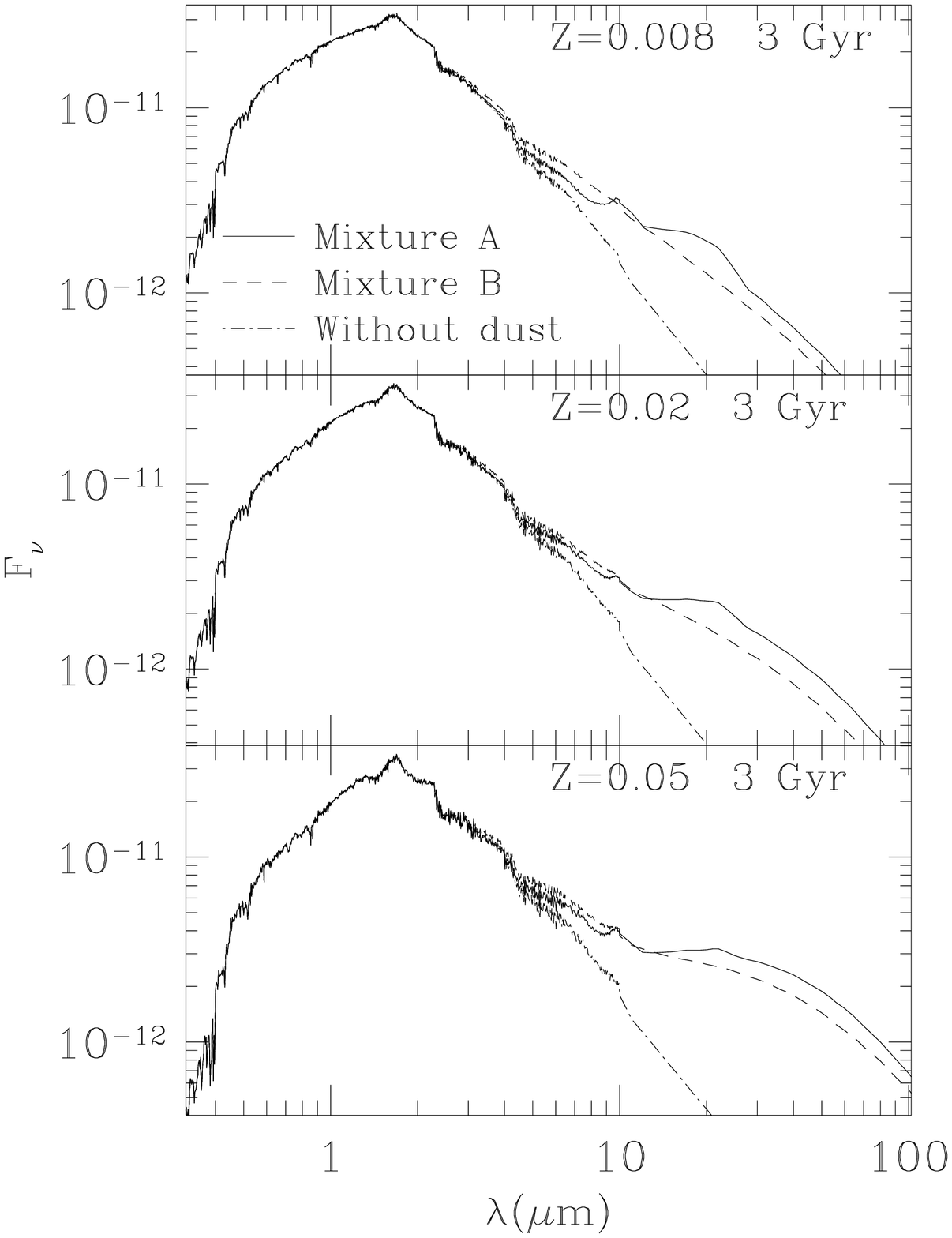,height=12truecm}
\caption{The same as in Fig. \ref{ssp02g} but for a fixed age of
3~Gyr and metallicity Z=0.008, Z=0.02 and Z=0.05, respectively.}
 \label{ssp02z}
 \end{figure}
\subsection{Integrated spectra}
The integrated spectra F$_\nu$ vs. $\lambda$ in $\mu$m of SSP of 0.2
1.5 and 5 Gyr are shown in Fig.\ \ref{ssp02g}. In each panel three
cases are depicted. In the standard SSP (dot-dashed line) the effects
of dusty circumstellar envelopes around Mira and OH/IR stars are
neglected. In the other two cases the circumstellar envelopes have
been included in modelling the AGB phase. The continuous line refers
to the mixture of silicate grains (mixture A) and it is well suited
for an AGB dominated by M giants, while the dashed line represents
models computed with the mixture of carbonaceous grains (mixture B)
and should be more appropriate for a population rich in C-stars. 
In the latter case the mid infrared spectra  are featureless as it is
expected from the optical properties of the carbonaceous grains (note
however that our mixture do not include PAH molecules, whose
possible formation in the photosphere of C-stars is discussed by
Helloing et
al. 1996).

Differences with respect to the standard SSP appear beyond a few
$\mu$m where the contribution of the brightest AGB stars to the
integrated light of the stellar population is large. After a
decrease of the flux in the region from 1 to 3 $\mu$m because the
brightest stars are heavily obscured by their circumstellar
envelopes, at larger wavelengths the contribution of the dust
dominates, reaching an order of magnitude at 10 $\mu$m over the pure
photospheric models. While there are clear differences between the
two selected dust mixtures, they constitute a modulation over the
main effect brought about by the inclusion of the dust in the
envelope. At increasing age the relative contribution of AGB stars to
the integrated light decreases and, correspondingly, the inclusion of
a proper circumstellar envelope becomes less and less important.

Metallicity effects are depicted in Fig.\ \ref{ssp02z} where we plot
SSPs of 1.5 Gyr for  Z=0.008, Z=0.02 and
Z=0.05 as indicated in the corresponding panels.
It can be noticed that the dust emission progressively shifts toward
larger wavelengths as the metal content increases, indicating that
the spectra are dominated by  cooler dust envelopes.
The metallicity affects the
properties of the dust envelope through Eq.\ \ref {eq_tau}.  By
combining this equation with Eq.\ \ref{mlr3}, describing the
super-wind phase, it is easy to show that the optical depth of the
envelope is a linear function of the metallicity Z and the maximum
value of $\tau$ attainable at a given metallicity scales as
L$^{-0.2}$.  Thus in the superwind phase, more metal rich isochrones
are characterized by a larger value of the optical depth. Moreover at
a given metallicity, older isochrones reach a larger value of the
maximum optical depth, but this effect is compensated by the lower
contribution of the AGB phase to the integrated light of the
population.

\subsection{Integrated colors}
\begin{figure*}
 \psfig{file=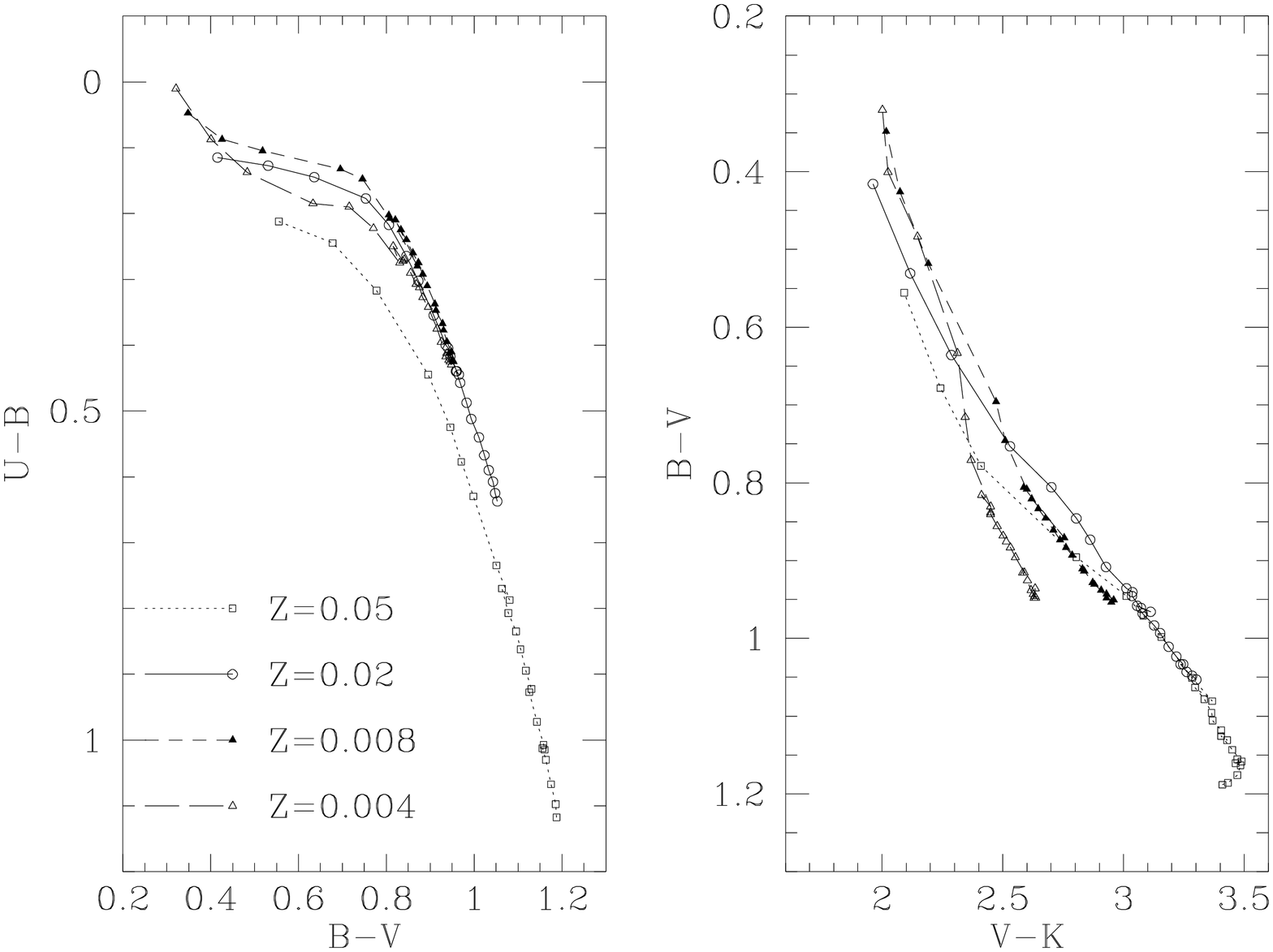,width=18truecm,angle=-90}
 \caption{Optical and near infrared two colors diagrams of SSPs at
 varying age and metal content illustrating the problem of the
 age-metallicity degeneracy. Age runs
from 0.5 Gyr (top left) to 15 Gyr (bottom right) in steps of 0.5 Gyr
up to 7 Gyr and thereafter in steps of 1 Gyr.}
 \label{opt_twoc}
 \end{figure*}

Accurate analysis of the integrated
properties of SSPs at optical wavelengths have been done in a
number of recent studies (Bertelli et al. 1994, Bressan, Chiosi \&
Fagotto 1994, Tantalo, Chiosi, Bressan \& Fagotto 1996, Charlot,
Worthey \& Bressan 1996). Figure \ref{opt_twoc} depicts optical
and near IR two-color plots of SSPs at different age and metallicity
and reveals one of the major problems encountered when working with
integrated properties: age and metallicity vectors are almost
superimposed in these diagrams and it is virtually impossible to
disentangle the two effects. In particular, because one of the
applications of the concept of SSP is concerned with the age
derivation of early type galaxies, this means that once a set of
observed colors or narrow band indices are fitted with models of a
given metallicity, there is always the possibility that either a
younger but more metal rich population or an older but more metal
poor one provides an equally acceptable fit. This effect has been
known for a long time  and it is the main source of uncertainty in
deriving absolute ages of early-type galaxies (Bressan, Chiosi \&
Fagotto 1994, Gonzales 1993 , Charlot, Worthey \& Bressan 1996).
Therefore finding a pair of indices whose age and metallicity
behavior is orthogonal, would allow a separation of these effects and
would provide an unambiguous estimate of both the age and metallicity
of a complex stellar system.
\begin{figure*}
 \psfig{file=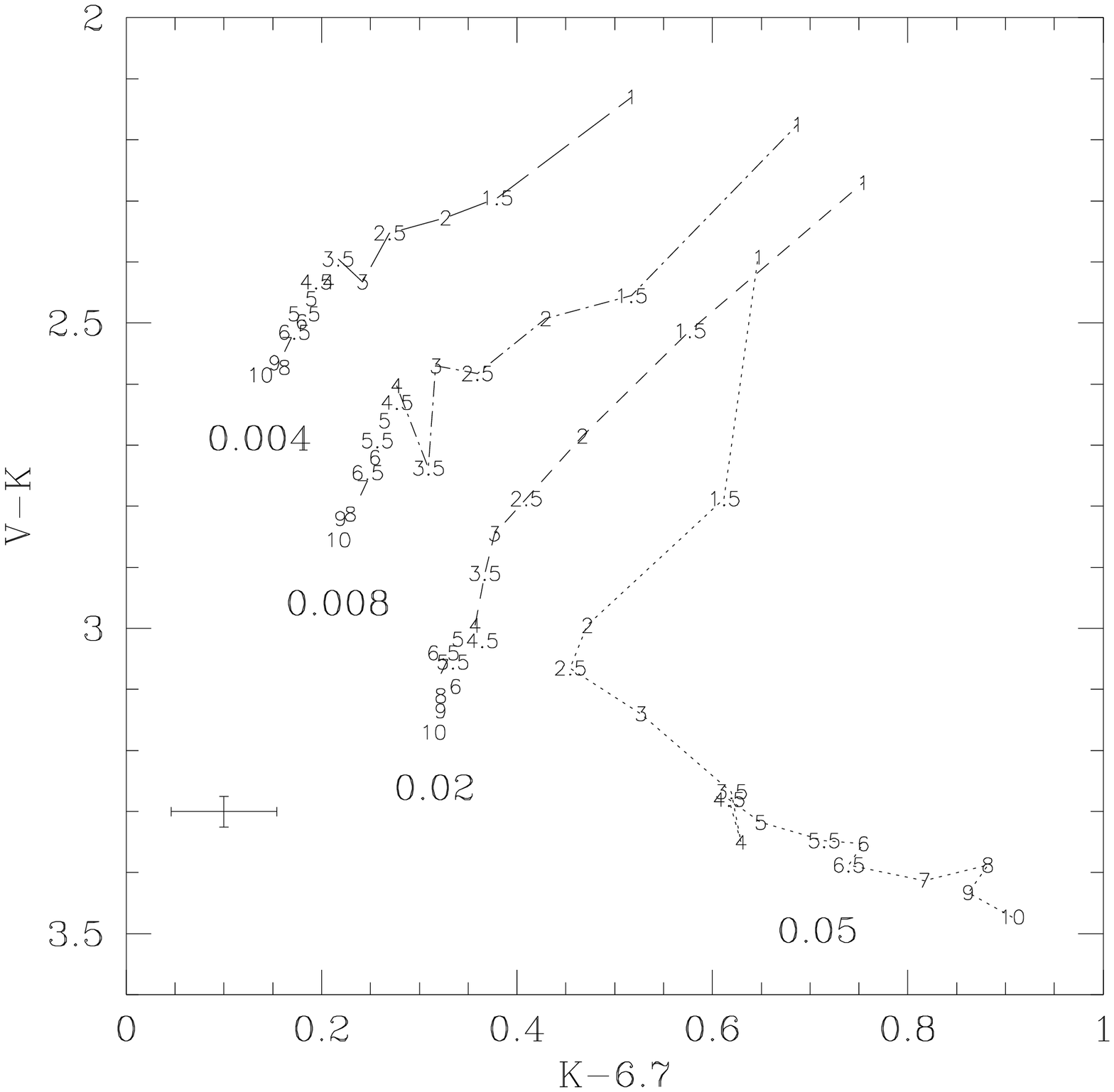,width=15cm}
 \caption{A mixed optical (V), near (K) and mid (6.7$\mu$m) infrared
 two-colors diagram of SSPs.  The metallicity Z is indicated in the
 figure. Numbers along the curves indicate the age in Gyr. The error
 bar corresponds to a 5\% and 10\% uncertainty in the flux
 measurement in the near and mid IR respectively. Due to dust effects
 in the envelope of AGB stars, age and metallicity differences may be
 fairly well recognized in intermediate age populations.}
 \label{vk_67}
 \end{figure*}
\begin{figure}
\psfig{file=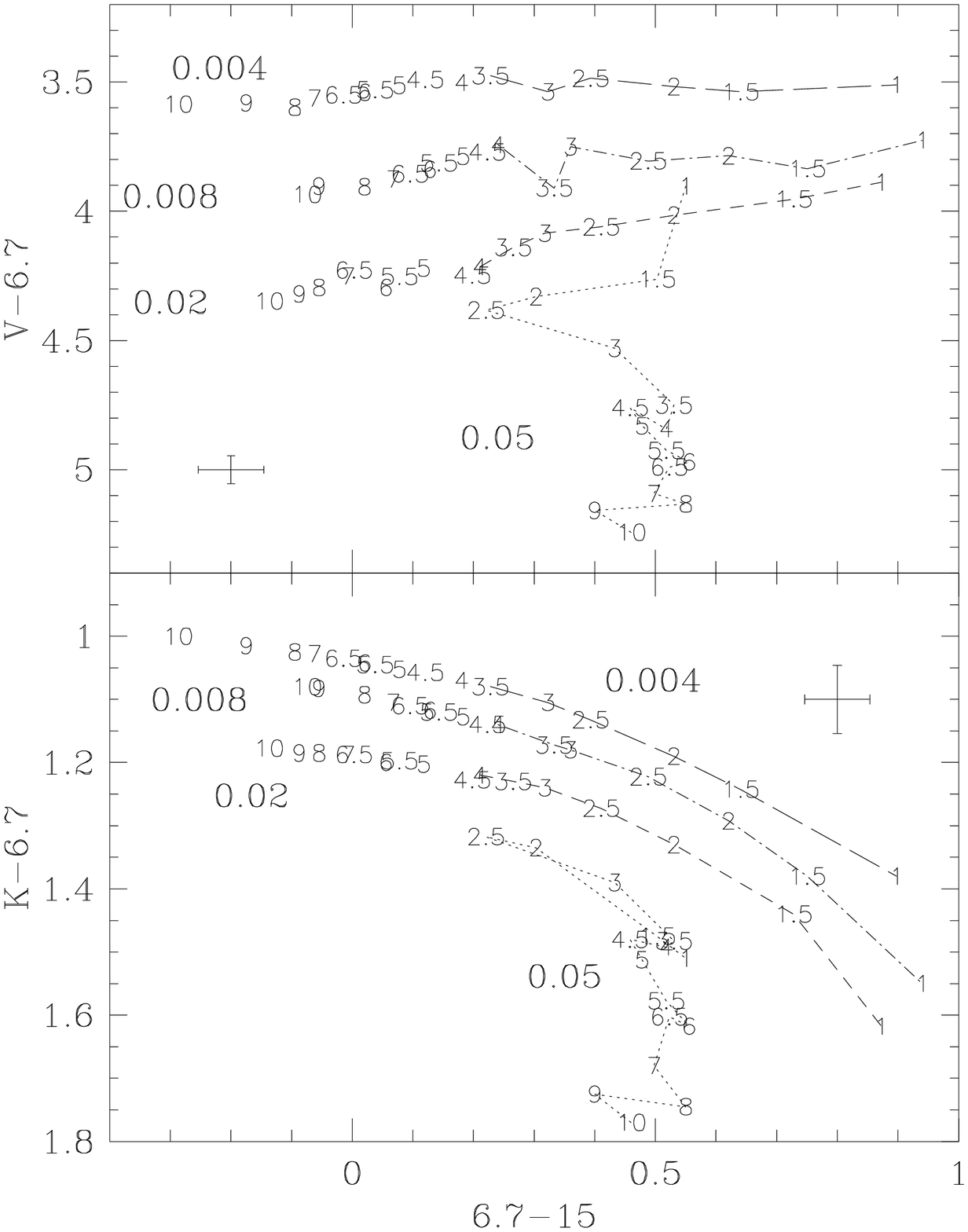,width=9truecm}
 \caption{The same as in Fig \ref{vk_67} but for the (V-6.7$\mu$m)
 and (6.7$\mu$m-15$\mu$m) colors, and (K-6.7$\mu$m) and
 (6.7$\mu$m-15$\mu$m) colors.}
 \label{ir}
 \end{figure}
Worthey et al.\ (1994) were able to show that one such pair is
provided by H$_\beta$ and MgFe indices. In the plane defined by
H$_\beta$ and MgFe indices the field elliptical galaxies analyzed by
Worthey et al.\ occupy a narrow band characterized by a small range
in MgFe (an index mainly related to the global metallicity) and a
range in H$_\beta$ corresponding to a significant dispersion in age
(from 15 to few Gyr). Bressan et al. (1996) however pointed out that
the H$_\beta$ index is very sensitive to recent episodes of star
formation that may have interested only a tiny fraction of the galaxy
mass, so that the age of the bulk of the population is still
inaccessible through this method, implying that our view of the
evolution of these objects still remains obscure. The degeneracy has
a major impact on the age determination of the high redshift
galaxies, since one of the major challenges offered by nowadays
observations is to look at galaxies when their age is only a fraction
of that of the universe. Bender et al. (1995) measured narrow band
indices of early-type galaxies in clusters down to redshift z
$\leq$0.3 finding that those systems are compatible with only a
passive evolution of their stellar populations. Recently Franceschini
et al.\ (1997a) combining HST magnitudes with JHK and ISO 6.7 $\mu$m
magnitude for a typical early-type galaxy in the Hubble Deep Field at
redshift z$\simeq$1 concluded that the bulk of the stellar population
is about 4 Gyr old and as massive as 4$\times$10$^{11}$ M$_\odot$. A
major episode of star formation must have occurred at redshift z = 1
+ $\Delta~z_{4 Gyr}$, which for a flat universe and H$_0$=50 km/s/Mpc
corresponds to z$\geq$5 and for a open $\Omega$=0 universe and
H$_0$=50 km/sec/Mpc to z$\geq$2.5. Our new SSP models show that
adopting a metal content larger by about the 50\%  would decrease the
age from  4 Gyr to about 2 Gyr, and the corresponding formation
redshift would be z$\leq$2 in both cases. While the obvious settling
of the problem is on a statistical basis, this example clearly
illustrates a feasible test for the models of galaxy formation.
Indeed, solar or even larger metallicities, as suggested by local
observations of narrow band indices and ultraviolet upturn (Bressan
et al. 1994, 1996) in elliptical galaxies and ages of few Gyr as
determined by fitting the broad band colors of intermediate redshift
objects, constitute the space of parameters where the AGB population
keeps its maximum effect upon the integrated star light and it is
possible that, because of the effect of the dust, the new SSP models
are most indicated for solving the age metallicity degeneracy,
provided that suitable pass-bands are selected. 

To illustrate the point,  Fig. \ref{vk_67} and Fig. \ref{ir} show
combined optical, near IR and mid IR two colors
diagrams of several SSPs computed adopting the silicate mixture. 
The 6.7$\mu$m and 15$\mu$m magnitudes are obtained by adopting a zero
point flux of 94.7 Jy and 17 Jy respectively and have been selected
for comparison with ISO data.
Fig. \ref{vk_67} depicts the run of the color (V-K) against the color
(K-6.7) for different metallicities (Z=0.004, 0.008, 0.02 and 0.05)
and some selected ages. Above 1.5 Gyr the different sequences are
well
separated, with leftmost sequence corresponding to the lowest metal
content and the rightmost one corresponding to Z=0.05. At
intermediate ages the effect of the metallicity is clearly at odd
with that of the age because AGB stars drive the mid IR fluxes of the
composite populations. Younger isochrones are  bluer in the (V-K)
color but redder in the (K-6.7) due to the effect of the dust, while
more metal rich isochrones are redder in both colors. This causes the
breaking of the age-metallicity degeneracy. Such a behavior can be
compared with that of optical and near infrared colors in order to
appreciate the power of the diagnostic tool we are suggesting.
Plotting the (V-K) color against (K-15$\mu$m) or (K-25$\mu$m) would
provide similar results, so that the corresponding figures are not
shown here for sake of conciseness.

The (V-6.7) vs. (6.7-15) two colors plot (upper panel of Fig.
\ref{ir}) is also a suitable diagnostic diagram. Note that there is a
variation of more than half a
magnitude in the (V-6.7) color going from Z=0.004 to Z=0.02 at
intermediate ages. In contrast, optical colors change by less than
0.1 mag in the same age range. Plotting an optical magnitude (e.g. V)
seems required in order to obtain useful diagnostic diagrams. Indeed
inspection of intermediate age SSPs show that while optical, near and
mid IR absolute magnitudes display the same trend with the age (they
increase by about one magnitude going from 2 to 6 Gyr, irrespective
of the metallicity), they behave differently with respect to the
metallicity. Optical magnitudes become fainter as the metallicity
increases, but the trend inverts when near IR bands are considered
and at 6.7$\mu$m, 15$\mu$m and 25$\mu$m there is a remarkable
brightening with the metallicity (dust makes the SSP more bright at
increasing metallicity). The lower panel of Fig. \ref{ir} contrasts
the (K-6.7) color instead of the (V-6.7) one. Disentangling age and
metallicity effects becomes now more difficult, because the fluxes in
the selected passbands show the same behavior with respect to the
metallicity.

While both the (V-K) vs. (K-6.7) and the (V-6.7) vs. (6.7-15) two
colors plots seem well suited for the purpose of disentangling age
and metallicity effects, some cautionary notes are necessary at this
point.

First of all we have {\it assumed} a linear dependence on the
metallicity of the fraction of the momentum carried out by radiation
responsible for the dust acceleration in the superwind phase. This
affects, in particular, $\tau_1$ for the most metal rich populations
and it is responsible of the odd behavior of the corresponding models
in the two colors plots. Such a relation should be tested against
forthcoming ISO observation of AGB stars.

Furthermore, SSPs with different dust mixture display different IR
spectra (Figs. \ref{ssp02g} and \ref{ssp02z}); in particular the SSPs
with the silicates mixture show features around 10-20 $\mu$m. Leaving
aside the problem of the dust composition in real cool stellar
envelopes, it is clear that the relative percentage of O-rich and
C-rich stars may affect the mid IR shape of the integrated spectrum.
When an O star becomes a C star (III dredge-up) and eventually turns
again into an O star (hot bottom burning), and how it depends on the
metallicity of the environment is still a matter of debate. We
neglected transient phenomena such as shell ejection or luminosity
deep, that may also affect the infrared colors of the star (e.g. Fig.
\ref{iras2}). Detailed modelling of such effects is underway (Marigo
et al., in preparation).

Some sort of dilution due to a range in age and metal content
should be present in the spectrum of a composite system, however this
could be of minor importance for early type galaxies if low
metallicity stars are practically absent (Bressan et al. 1994) and
the formation time is short, as suggested by the observed enhanced
composition (Bressan et al. 1996).

Finally, environmental effects should be considered
when adopting these models to interpret the colors of galaxies.
The presence of a low level star formation activity could mask the
infrared AGB emission. But the opposite is also true: the models
indicate that the mid IR emission of an intermediate age population
alone is up to an order of magnitude larger than predicted by pure
photospheric models. Also the presence of a diffuse gas phase, the
cirrus component, could dominate the galactic spectrum beyond 20
$\mu$m. All these phenomena are the subject of a forthcoming
investigation (Silva et al. 1997).

\section{Conclusions}
In this paper we modelled the infrared emission of late type stars
surrounded by a steady-state outflows of matter driven by radiation
pressure on dust grains. To this aim we solved the radiative
transport equations inside a spherically symmetric stationary flow of
matter, in order to derive the extinction and the emission of the
dust as a function of the wavelength. Several envelope models were
constructed as a function of the optical depth of the envelope
$\tau_{1}$, a suitable scaling parameter for the models (see also
Ivezic \& Elitzur 1995). We also found a relation between this
scaling parameter $\tau_{1}$ and the mass-loss rate, expansion
velocity, luminosity and metallicity. Finally by adopting some
empirical relations we could express this quantity as a function of
the basic stellar parameters, mass, luminosity effective temperature
and metal content, in such a way that we could assign the suitable
dust envelope to the stars along the AGB.

We thus computed isochrones at different ages and metallicity where,
for the first time, the effect of the dust envelope around the
brightest AGB stars is accounted for in a consistent way. Comparing
our isochrones with the sample of IRAS sources defined by Van der
Veen \& Habing (1988) in the two colors diagram we run into the
problem of the dispersion of the data perpendicularly to the sequence
defined by different value of $\tau_1$. We briefly discuss the
possibility of overcome this difficulty by adopting a different grain
mixture or by allowing the occurrence of a density enhancement caused
by an outward expanding shell. While both these effects can account
for the mentioned dispersion, we considered the ability of our
stationary models to fit the bulk of the data in the IRAS two-color
diagram as a good test of reliability, and adopted these models to
compute the integrated properties of the SSP.

As a further test we compared our isochrones with a sample of
galactic M and Mira stars from Whitelock et al. (1994, 1995). In the
near infrared colors (J-H vs. H-K) the models fit nicely both the M
giants and the Miras.  Models without the effect of dust cannot fit
the location of the Mira stars, as indicated by the location in this
diagram of our adopted model atmospheres. Apparently the sample of
Mira stars is best fitted by models with solar metallicity, however
this conclusion is much dependent on the adopted temperature scale of
the more advanced M spectral types other than by the atmospheric
model itself. Forthcoming ISO observations would certainly help in
clarifying this point.

Existing studies based on theoretical SSP either do not account for
the effect of dust in the envelope of bright AGB stars or they simply
give a rough estimate of the effect and, consequently, neither class
of models is fully suitable for investigating the integrated
properties of stellar systems in the infrared (Charlot, Worthey \&
Bressan 1996). In our present calculations the mass-loss along the
AGB determines both the lifetime and the spectral properties of the
stars, allowing the description of the whole phase inside a
consistent framework.

The direct effect of the metallicity is included by a simple
parameterization of the superwind mass-loss rate, which finally
provides the scaling parameter $\tau_{1}$. However the metal content
also affects the effective temperature of the AGB phase and
consequently the mass-loss rate determined by the empirical relation
with the pulsational period.

Effects of AGB stars show up even in the integrated spectrum of the
population, at wavelength above few $\mu$m where they are the largest
contributors to the integrated light. At these wavelengths integrated
spectra computed without the effect of dust envelopes are about one
order of magnitude fainter than the models including Miras and OH/IR
stars. This effect decreases as the SSP becomes older and older.

We show that by selecting suitable passbands, the integrated colors
of such new SSPs can potentially be used to disentangle age from
metallicity effects because of the strong dependence of the dust
model on the metal content. However given the short duration and the
high luminosity of the phase this method will suffer of a great
stochastic effect when applied to systems with a relatively small
number of stars (Chiosi et al. 1988, Santos \& Frogel 1997). These
stochastic effects would certainly constitute less of a problem in
the interpretation of the integrated light of large systems such as
the nuclei of elliptical galaxies or the spiral bulges. These
considerations suggest that incoming infrared observations of
intermediate redshift objects could be the targets where this method
will eventually exploit its maximum capability.

Finally, the new SSP models will be included in a more extended study
aimed at modelling the spectral properties of the galaxies from the
ultraviolet to the far infrared, accounting for the effects of the
dust associated with star forming regions and with the diffuse
interstellar gas (Silva et al. 1997, Franceschini et al.\ 1997b)

\acknowledgements{We thank Paola Marigo, Alberto Franceschini and Leo
Girardi for helpful discussions. We also thank our referee, A.
Lan\c{c}on for her useful comments and suggestions. This research was
supported by the European Community under TMR grant ERBFMRX-CT96-0086}

\section*{References}
Bender R., Ziegler B., Bruzual G., 1996 \APJ 463 51.
\protect\\
Bertelli G., Bressan A., Chiosi C., Fagotto F., Nasi E., 1994, \AAS
106, 275
\protect\\
Bedijn P. J. 1987 \AAP 186 136.
\protect\\
Bertola, F., Bressan, A., Burstein, D., Buson, L.M., Chiosi, C.,
Di Serego Alighieri, S., 1995: \APJ 438 680.
\protect\\
Bessell M.S., Wood P.R., Brett J.M., Scholtz M., 1991 \AAS 89 335.
\protect\\
Bl\"ocker, T., 1995a: \AAP 297 727.
\protect\\
Bl\"ocker, T., 1995b: \AAP 299 755.
\protect\\
Bl\"ocker, T., Sch\"onberner, D., 1991: \AAL 244 43.
\protect\\
Bowen, G.H., Willson, L.A., 1991: \APJL 375 53
\protect\\
Bressan, A., Fagotto, F., Bertelli, G., Chiosi, C., 1993, \AAS 100
647.
\protect\\
Bressan, A., Chiosi, C., Fagotto, F., 1994: \APJS 94 63.
\protect\\
Bressan, A., Chiosi, C., Tantalo, R., 1996 \AAP 311 425
\protect\\
Bressan, A., Chiosi, C., 1996: in preparation.
\protect\\
Bruzual, A.G., Charlot, S., 1993: \APJ 273 205.
\protect\\
Burstein, D., Bertola, F., Buson, L.M., Faber, S.M., Lauer, T.R.,
1988: \APJ
328 440.
\protect\\
Carraro, G., Girardi, L., Bressan, A., Chiosi, C., 1996: \AAP 305
849.
\protect\\
Castor J.J, Abbot D.C., Klein R.I., 1975 \APJ 195 157.
\protect\\
Charlot, S., Worthey, G., Bressan, A., 1996: \APJ 457 625.
\protect\\
Chiosi, C., Bertelli, G., Bressan, A. 1988, \AAP 196 84
\protect\\
Collison, A.J., Fix, J.D., 1991 \APJ 368 545.
\protect\\
Elitzur, M., Goldreich, P., Scoville, N., 1976 \APJ 205 384.
\protect\\
Fagotto, F., Bressan, A., Bertelli, G., Chiosi, C., 1994a \AAP 104
365.
\protect\\
Fagotto, F., Bressan, A., Bertelli, G., Chiosi, C., 1994b \AAP 105
369.
\protect\\
Fluks M.A., Plez B., The P.S., De Winter D., Westerlund B.E.,
Steenman H.C.,
1994, \AAS 105 311.
\protect\\
Franceschini, A. et al.\ 1997a in ESA SP-401.
\protect\\
Franceschini, A., Silva, L., Granato, G.L.,  Bressan, A., Danese, L.,
1997b {\em Astrophys. J.}, submitted.
\protect\\
Freedman, W., 1992: \AJ 104 1349.
\protect\\
Fusi Pecci, F., Renzini, A.  1976, \AAP 46 447.
\protect\\
Goldreich, P., Scoville, N., 1976 \APJ 205 144.
\protect\\
Gonzales J.J. 1993, Ph. D. thesis, Univ. California, Santa Cruz.
\protect\\
Granato G.L., Danese L., 1994 \MN 268 235.
\protect\\
Groenewegen M.A.T. \& De Jong T., 1994 \AAP 283 463.
\protect\\
Groenewegen M.A.T., Smith C.H., Wood P.R., Omont A., Fujiyoshi, 1995:
\APJL 449
119.
\protect\\
Habing, H.J.,1996, \AAR 7 97.
\protect\\
Habing, H.J., Tignon, J., Tielens A.G.G.M., 1994, \AAP 286 523.
\protect\\
Helling, C., Jorgensen, U.G., Pletz, B., Johnson, H.R., 1996, \AAP
315 194.
\protect\\
Iben, I.J., 1984, \APJ 277 333.
\protect\\
IRAS Science Team, 1988, IRAS PSC explanatory supplement.
\protect\\
Ivezic Z., Elitzur M., 1995 \APJ 445 415
\protect\\
Justtanont K., Tielens A.G.G.M., 1992 \APJ 389 400.
\protect\\
Kent, 1987: \AJ 94 306.
\protect\\
Kurucz R.L., 1992 private communication.
\protect\\
Marigo P., Bressan A., Chiosi C., 1996 \AAP 313 545
\protect\\
Marigo P., Bressan A., Chiosi C., 1997 A\&A submitted
\protect\\
Montgomery, K.A., Marschall, L.A., Janes, K.A., 1993: \AJ 106 181.
\protect\\
Netzer N., Elitzur M., 1993 \APJ 410 701
\protect\\
O'Connel, R., 1986: in Spectral Evolution of Galaxies, ed. Chiosi \&
Renzini
(Dordecht: Reidel), 195.
\protect\\
Paczy\'nski, B., 1971: \ACTA 21 417.
\protect\\
Renzini, A., Buzzoni, A, 1986: in Spectral Evolution of Galaxies, ed.
Chiosi \&
Renzini (Dordecht: Reidel), 195.
\protect\\
Rowan-Robinson, M., 1986 \MN 219 737.
\protect\\
Salpeter, E.E., 1974a, \APJ 193 579
\protect\\
Salpeter, E.E., 1974b, \APJ 193 585
\protect\\
Santos, J. F. C. Jr. \& Frogel Jay A., 1997, \APJ 479 764
\protect\\
Sch\"onberner, D., 1981 \AAP 103 119.
\protect\\
Sch\"onberner, D., 1983 \APJ 272 708.
\protect\\
Silva L., 1995, Degree thesis, Univ. Padova, IT
\protect\\
Silva, L., Granato, G.L., Bressan, A. \& Danese, L., 1997 {\em
Astrophys. J.},
submitted. 
\protect\\
Tantalo, R., Chiosi, C., Bressan, A., Fagotto, F., 1996, \AAP 311
361.
\protect\\
Tripicco, M.J., Dorman, B., Bell, R.A., 1993 \AJ 106 618.
\protect\\
Van Der Veen W.E.C.J., Habing H.J., 1988, \AAP 194 125
\protect\\
Vassiliadis E. \& Wood P.R., 1993: \APJ 413 641.
\protect\\
Weidemann, V., 1987: \AAP 188 74.
\protect\\
Willson L.A., Bowen G.H., Struck C.,1995 BAAS 187 103.18
\protect\\
Whitelock P., Menzies J., Feast M., Marang F., Carter B., Roberts G.,
Catchpole R., Chapman J., 1994, \MN 267 711.
\protect\\
Whitelock P., Menzies J., Feast M., Catchpole R., Marang F., Carter
B., 1995,
\MN 276 219.
\protect\\
Wood P.R., Whiteoak J. B., Hughes S. M. G, Bessel M. S., Gardner F.
F. \&
Hyland A. R., 1992: \APJ 397 552.
\protect\\
Worthey G., Faber S.M., Gonzalez J.J., Burstein D., 1994, \APJS 94
687.

\end{document}